\documentclass[iop]{emulateapj}
\pdfoutput=1
\usepackage{graphicx}

\newcommand{\eg}{e.g.,\ }

\newcommand{\etal}{et~al.\ }
\newcommand{\ltsima}{$\; \buildrel < \over \sim \;$}
\newcommand{\simlt}{\lower.5ex\hbox{\ltsima}}
\newcommand{\gtsima}{$\; \buildrel > \over \sim \;$}
\newcommand{\simgt}{\lower.5ex\hbox{\gtsima}}
\newcommand{\kms}{km~s$^{-1}$}

\def\mub{$\mu_B$}

\def\R25{R$_{25}$}
\def\D25{D$_{25}$}

\def\Msun{M$_\sun$}
\def\Lsun{L$_\sun$}
\def\MHI{M$_{\rm HI}$}
\def\vlsr{V$_{\rm LSR}$}
\def\W20{$W_{20}$}

\begin{document}

\title{The HI environment of the M101 group}

\author{J. Christopher Mihos,\altaffilmark{1}
	Katie M. Keating,\altaffilmark{2}
      Kelly Holley-Bockelmann,\altaffilmark{3,4}\\
     D.J. Pisano,\altaffilmark{5,6} and
    Namir E. Kassim\altaffilmark{7}}
\email{mihos@case.edu, kmk@rincon.com, k.holley@vanderbilt.edu,
        djpisano@mail.wvu.edu, namir.kassim@nrl.navy.mil}

\altaffiltext{1}{Department of Astronomy, Case Western Reserve University,
Cleveland, OH 44106}

\altaffiltext{2}{Rincon Research Corporation, Tucson, AZ 85711}

\altaffiltext{3}{Department of Physics and Astronomy, Vanderbilt University, Nashville, TN 37235}

\altaffiltext{4}{Department of Physics, Fisk University, Nashville, TN 37208}

\altaffiltext{5}{Department of Physics, West Virginia University, Morgantown, WV 26506}

\altaffiltext{6}{Adjunct Assistant Astronomer at National Radio Astronomy
Observatory}

\altaffiltext{7}{Naval Research Laboratory, Washington, DC 20375}

\begin{abstract}

We present a wide (8.5$^{\circ}$ $\times$ 6.7$^{\circ}$, 1050 $\times$
825 kpc), deep ($\sigma_{\rm N_{HI}}$= 10$^{16.8-17.5}$ cm$^{-2}$)
neutral hydrogen (HI) map of the M101 galaxy group. We identify two new
HI sources in the group environment, one an extremely low surface
brightness (and hitherto unknown) dwarf galaxy, and the other a starless
HI cloud, possibly primordial in origin. Our data show that M101's
extended HI envelope takes the form of a
$\sim$ 100 kpc long tidal loop or plume of HI extending to the southwest
of the galaxy. The plume has an HI mass $\sim$ 10$^8$ \Msun\ and a peak
column density of N$_{\rm HI}$= 5 $\times$ 10$^{17}$ cm$^{-2}$, and
while it rotates with the main body of M101, it shows kinematic
peculiarities suggestive of a warp or flaring out of the rotation plane
of the galaxy. We also find two new HI clouds near the plume with masses
$\sim$ 10$^7$ \Msun\llap, similar to HI clouds seen in the M81/M82
group, and likely also tidal in nature. Comparing to deep optical
imaging of the M101 group, neither the plume nor the clouds have any
extended optical counterparts down to a limiting surface brightness of
\mub = 29.5. We also trace HI at intermediate velocities between M101
and NGC 5474, strengthening the case for a recent interaction between
the two galaxies. The kinematically complex HI structure in the M101
group, coupled with the optical morphology of M101 and its companions,
suggests that the group is in a dynamically active state that is likely
common for galaxies in group environments.

\end{abstract}

\keywords{galaxies:dwarf -- galaxies:evolution -- galaxies:groups -- 
galaxies:individual:M101 -- galaxies:interactions -- galaxies:ISM}

\section{Introduction}

The gaseous ecosystems around nearby galaxies are thought to be rich and
dynamic, reflecting a wide variety of processes associated with galaxy
formation and evolution. Theoretical arguments predict that galaxies can
continually accrete gas from their surroundings, as material falls in
from the more diffuse intragroup medium (\eg Larson 1972; Maller \&
Bullock 2004; Kere\v s \etal 2005; Sommer-Larsen 2006). Massive galaxies
can host a significant population of gas-rich dwarf companions, and
interactions with these companions and with larger nearby galaxies can
imprint kinematic and morphological signatures in the extended HI disks,
such as disk warps, HI plumes and tidal tails, and high velocity cloud
complexes (see compilations in Hibbard \etal 2001; Sancisi \etal 2008).
Radially extended gas is responsive to even low-level perturbations, and
this makes deep HI imaging an excellent tool to pinpoint subtle
dynamical effects that are missed in the optical, and that help drive
galaxy evolution.

Galaxies are believed to be embedded in a gaseous network which can feed
galaxies through accretion. At high redshift, this accretion is thought
to be rapid, perhaps in a ``cold" accretion mode where gas flows
smoothly onto galaxies along dense dark matter filaments, never shock
heating to the virial temperature of the host dark matter halo (\eg Kere\v s \etal 
2005, 2009, Dekel \& Birnboim 2006, Stewart \etal 2011).
However, under these models, the fraction of cold gas accretion plummets
once the dark matter halo reaches a critical mass $\sim 10^{12}$ \Msun
(Dekel \& Birnboim 2006). At the present epoch, therefore, the accretion
of gas onto massive galaxies may happen either via ``hot mode"
accretion, where gas cools from a surrounding hot halo, or by the
accretion of smaller gas-rich companion galaxies. Of course, these
scenarios are not mutually exclusive, as several accretion processes may
be operating simultaneously. The exact mechanism driving gas accretion
onto galaxies at low redshift --- whether it be hot mode, vestigial cold
mode, or satellite accretion --- remains unclear.

Searches for signatures of gas accretion have shown a very complex HI
environment around nearby galaxies. On large scales, HI tails, plumes,
warps, and lopsidedness certainly argue that ongoing accretion is
triggered by interactions and mergers (see Sancisi \etal 2008 for a
review). In many cases, the clear culprit is a companion galaxy tidally
stripping and/or being stripped the host galaxy. In contrast, searches
for pure ``starless" HI clouds in the intragroup medium that might drive
gaseous accretion have yielded few detections (\eg de Blok \etal 2002;
Kova\v c \etal 2009; Chynoweth \etal 2009; Pisano
\etal 2007, 2011). While strongly interacting groups can host a
population of massive free-floating HI clouds (such as those found in
the M81/M82 group; Chynoweth \etal 2008), the clouds are typically
linked to the more diffuse tidal debris and have a phase space structure
consistent with a tidal formation scenario, rather than being fresh
material accreting from the intragroup medium or being ``dark galaxies''
hosted by their own dark matter halo (Chynoweth \etal 2011a). In short,
despite solid theoretical evidence that galaxies should accrete gas from
the intergalactic medium, or draw from the warm gas reservoir in the
halo, most of the observed HI structures are consistent with interaction
with or accretion from another gas-rich galaxy.

The ubiquity of tidal interaction signatures in the HI distribution
around nearby galaxies is easily understood by the radially extended
nature of the neutral interstellar medium in galaxies. Because gas at large radius is
less tightly bound to its host galaxy than is the more concentrated
stellar component, it is {\it more} responsive to tidal perturbations,
and can reveal interaction signatures even in systems which show little
or no direct evidence of tidal debris in the optical, such as the Leo
Ring around the M96 group (Schneider 1985; Michel-Dansac \etal
2010), the tidal features connecting galaxies in the M81/M82 system
(Yun \etal 1994), or the tidal tail in the Leo Triplet, first
detected in HI (Haynes \etal 1979). Beyond simply
identifying interacting systems, the morphology and kinematics of the HI
tidal debris can be used to develop detailed dynamical models of the
interaction (\eg Hibbard \& Mihos 1995; Yun 1999; Michel-Dansac \etal
2010; Barnes 2011)

The nearby\footnote{In this paper, we adopt a physical distance to M101
of 6.9 Mpc, but note that the exact distance to M101 remains uncertain,
with recent estimates ranging from 6.1 to 7.6 Mpc (see, \eg the recent
compilation by Matheson \etal 2012).} M101 galaxy group (Tully 1988)
provides an opportunity to study the connection between the group HI
environment and the properties of the individual galaxies. M101 itself
is a well-studied, massive Sc galaxy, and its highly asymmetric disk
suggests ongoing interactions with other galaxies within the group
(Beale \& Davies 1969; Waller \etal 1997). Previous 21-cm neutral atomic
hydrogen (HI) imaging of M101, both single dish and synthesis mapping,
has shown a complex HI environment: a distorted HI disk with several
high velocity cloud complexes (van der Hulst \& Sancisi 1988, hereafter vdHS88;
Walter \etal 2008), and an extended, asymmetric outer HI plume at lower
column density (Huchtmeier \& Witzel 1979, hereafter HW79]). The outer disk shows
blue optical colors (Mihos \etal 2012) and extended UV emission (Thilker
\etal 2007), consistent with extended star formation, perhaps triggered
by a recent accretion or interaction event.

In this paper we survey the neutral hydrogen content of the M101 group
using the 100m Robert C. Byrd Green Bank Telescope (GBT) at the
NRAO\footnote{The National Radio Astronomy Observatory (NRAO) is a
facility of the National Science Foundation operated under cooperative
agreement by Associated Universities, Inc.} in Green Bank, West
Virginia. Compared with previous single-dish 21-cm studies of M101 (HW77),
our survey is wider in area, more sensitive, and covers a wider velocity
range. Our data is complementary to HI synthesis maps of galaxies in the
M101 group by probing the more extended, low column density HI
surrounding the galaxies. We use the data to search for isolated HI
clouds in the M101 group, to identify any signatures of HI accretion
onto M101, to probe the kinematics of the diffuse HI around M101, and to
search for any extended HI that might trace a past interaction between
M101 and any of its companions.

\section{Observations}

We observed the M101 galaxy group using the GBT auto-correlation
spectrometer using 9-bit sampling. The observations were carried out in
20 sessions between 2011 January and April. We observed the inner 2$^{\circ}$
$\times$ 1.5$^{\circ}$ region surrounding M101 more sensitively than the
outer regions of the galaxy group, in order to search more deeply for HI
features. To survey the entire M101 group, we moved the telescope in a
``basket-weave" pattern with 3$\arcmin$ offsets between each scan and
sampled every 3$\arcmin$ at an integration time of 2 s per sample.
This sampling interval corresponds to a slightly better-than-Nyquist
rate at approximately 3 pixels per beam. Total integration time was
approximately 86 hr. Table \ref{m101obs} gives a summary of the
observations, including the rms noise figures.

We calibrated the GBT data in the standard manner using the GBTIDL and
AIPS\footnote{Developed by the National Radio Astronomy Observatory;
documentation at http://gbtidl.sourceforge.net,
http://www.aoc.nrao.edu/aips} data reduction packages. A short
observation of a discrete reference position outside the map of the
galaxy group bracketed 1--4 hr sections of integration time.
We averaged the reference position observations bracketing each section
of data to construct an average reference spectrum for that section, and
used the reference spectrum to perform a standard
(signal-reference)/reference calibration of each pixel. Spectra were
then smoothed from their native resolution of 3.1 kHz to a channel
spacing of 24.4 kHz, corresponding to a velocity resolution of 5.2 km
s$^{-1}$ across the 12.5 MHz bandwidth. This spectral smoothing balances
the need for sensitivity, which is increased with more spectral
smoothing, with the need to resolve possible narrow line-width HI
features. The calibrated spectra were scaled by the system temperature
(typically about 20 K for these observations), corrected for atmospheric
opacity and GBT efficiency. We adopted the GBT efficiency Equation (1)
from Langston \& Turner (2007) with a zenith atmospheric opacity of
$\tau_{0}$ = 0.009 and a zenith efficiency of $\eta_A$ = 0.71.

The frequency range observed was relatively free of radio-frequency
interference (RFI), with less than 0.5\% of all spectra adversely
affected. The spectra exhibiting RFI were identified by tabulating the
rms noise level in channels free of neutral hydrogen emission. Spectra
that showed high rms noise across many channels were flagged and
removed. Narrow-band RFI also adversely affected a small number of
channels, which were blanked from the datacube. The data were then
gridded using the AIPS task SDIMG, which also averages polarizations. We
used a Gaussian-tapered Bessel convolving function for gridding as
described in Mangum et al. (2007), with parameters found empirically to
maximize the map sensitivity without degrading resolution (F. J.
Lockman, private communication). After amplitude calibration and
gridding, continuum sources were subtracted by fitting and subtracting a
first-order polynomial to line-free regions of the spectra. Only channels
in high negative velocity ranges, where few or no sources are expected,
were used for the fit. This simple baseline fit was extrapolated to the
positive velocity range, where galaxies and unknown HI features make a
baseline fit unreliable.

The effective angular resolution, determined from maps of 3C286, is
9.15\arcmin $\pm$ 0.05\arcmin -- corresponding to a physical resolution
of 18.7 kpc. To convert to units of flux density, we observed the
calibration source 3C286, whose flux density is 14.57 $\pm$ 0.94 Jy at
1.418 GHz (Ott \etal 1994). The calibration from K to Jy was derived by
mapping 3C286 in the same way as the HI maps were produced. After all
corrections for the GBT efficiency and the mapping process, the scale
factor from K beam$^{-1}$ to Jy beam$^{-1}$ images is 0.49.

The rms uncertainty varies across the field due to differing integration
times; in the inner region around M101 itself, the rms noise is $\sim$ 3
mJy beam$^{-1}$, while in the outer region the noise is a factor of five
larger, $\sim$ 15 mJy beam$^{-1}$. These noise levels correspond to a
limiting column density for the dataset of $\log({\rm N_{HI}})$ = 16.8
and 17.5 in the inner and outer regions, respectively. To calculate a
characteristic HI mass sensitivity, we assume that low mass clouds will be
unresolved in our beam, and use the relation

\begin{equation}
\left(\frac{\sigma_{M}}{M_{\odot}}\right)=2.36 \times 10^5 \left(\frac{D}{{\rm Mpc}}\right)^2 \left(\frac{\sigma_{s}}{{\rm Jy} }\right)
\left(\frac{\Delta V}{{\rm km}\hskip 1 mm {\rm s}^{-1}}\right),
\end{equation}

\noindent where $D$ is the distance to M101 in Mpc, $\sigma_s$ is the
rms noise in one channel, and $\Delta V$ is the channel width. For our
adopted distance of $D=6.9$ Mpc, we obtain a $1\sigma$ sensitivity per
5.2 \kms\ channel of $1.75\times 10^5$ \Msun\ in the inner region of our
map and $8.75\times 10^5$ \Msun\ in the outer region. However, in
searching for discrete HI emission, we used more conservative detection
limits (described below) to ensure valid detections.

To systematically survey the M101 group for HI emission, we extract a
spectrum for each spatial pixel, masking out velocity channels in the
range \vlsr$=-171.4 {\rm\ to\ } +45.0$ \kms, which are dominated by
diffuse Galactic emission, and channels in the range \vlsr$=+91.4 {\rm\
to\ } +117.1$ \kms, which were contaminated by narrowband RFI. In each
spectrum, we then search for emission that is $2.5\sigma$ above the
noise level in three adjacent velocity channels, with at least one
channel having peak emission of $3.5\sigma$. Objects detected this way
are thus detected at a $5\sigma$ level integrated over three
channels. To determine the limiting detectable cloud mass, we insert
artificial point source clouds of varying HI mass and gaussian velocity
widths into the datacube and see how effectively our search algorithm
recovers these clouds. At \W20=30 \kms, our 90(50)\% recovery limit is
2(1.6)$\times 10^6$ \Msun\ in the inner region and 10(7.5)$\times 10^6$
\Msun\ in the outer region. Larger velocity widths spread the emission
out over more channels and raises our mass detection threshold; at
\W20=60 \kms, the 90(50)\% limits rise to 3.5(2.5)$\times 10^6$ \Msun\
in the inner region and 16(12)$\times 10^6$ \Msun\ in the outer region.

\begin{table}
\begin{center}
\caption{M101 Group Observations Summary}
\label{m101obs}
\begin{tabular}{ll}
\tableline\tableline
Area: & \\
\hskip 3mm $\alpha$ range (J2000):  \dotfill &  13:26:39.1 -- 14:30:27.7 \\
\hskip 3mm $\delta$ range (J2000): \dotfill &  50:46:13 -- 57:50:27  \\
Observations: & \\
\hskip 3mm Center frequency (MHz): \dotfill &  1418 \\
\hskip 3mm Bandwidth (MHz): \dotfill & 12.5 \\
\hskip 3mm Velocity range (heliocentric, km s$^{-1}$): \dotfill & --787 -- +1855 \\
\hskip 3mm Channel width (kHz): \dotfill & 24.4 \\
\hskip 3mm Velocity resolution (km s$^{-1)}$ \dotfill & 5.2 \\
Integration time (hours): \dotfill & 86 \\
Sensitivity, inner region: & \\
\hskip 3mm rms noise (mJy beam$^{-1}$) \dotfill & 3.0 \\
\hskip 3mm N$_{\rm HI}$ (cm$^{-2}$) \dotfill & 10$^{16.8}$ \\
Sensitivity, outer region: & \\
\hskip 3mm rms noise (mJy beam$^{-1}$) \dotfill & 15.0 \\
\hskip 3mm N$_{\rm HI}$ (cm$^{-2}$) \dotfill & 10$^{17.5}$ \\
Beam size: & \\
\hskip 3mm Angular ($\arcmin$): \dotfill & 9.1 \\
\hskip 3mm Physical (kpc): \dotfill & 18.2  \\
\tableline
\end{tabular}
\tablecomments{Sensitivities are $1\sigma$ per 5.2 \kms\ velocity channel.}
\end{center}
\end{table}

\section{Results}

\subsection{The HI Environment of the M101 Group}

We start our analysis with a survey of the HI environment on large
scales around M101; a detailed discussion of the HI distribution and
kinematics of M101 itself follows in Section 3.2.

\begin{figure*}[]
\centerline{\includegraphics[width=6.0truein]{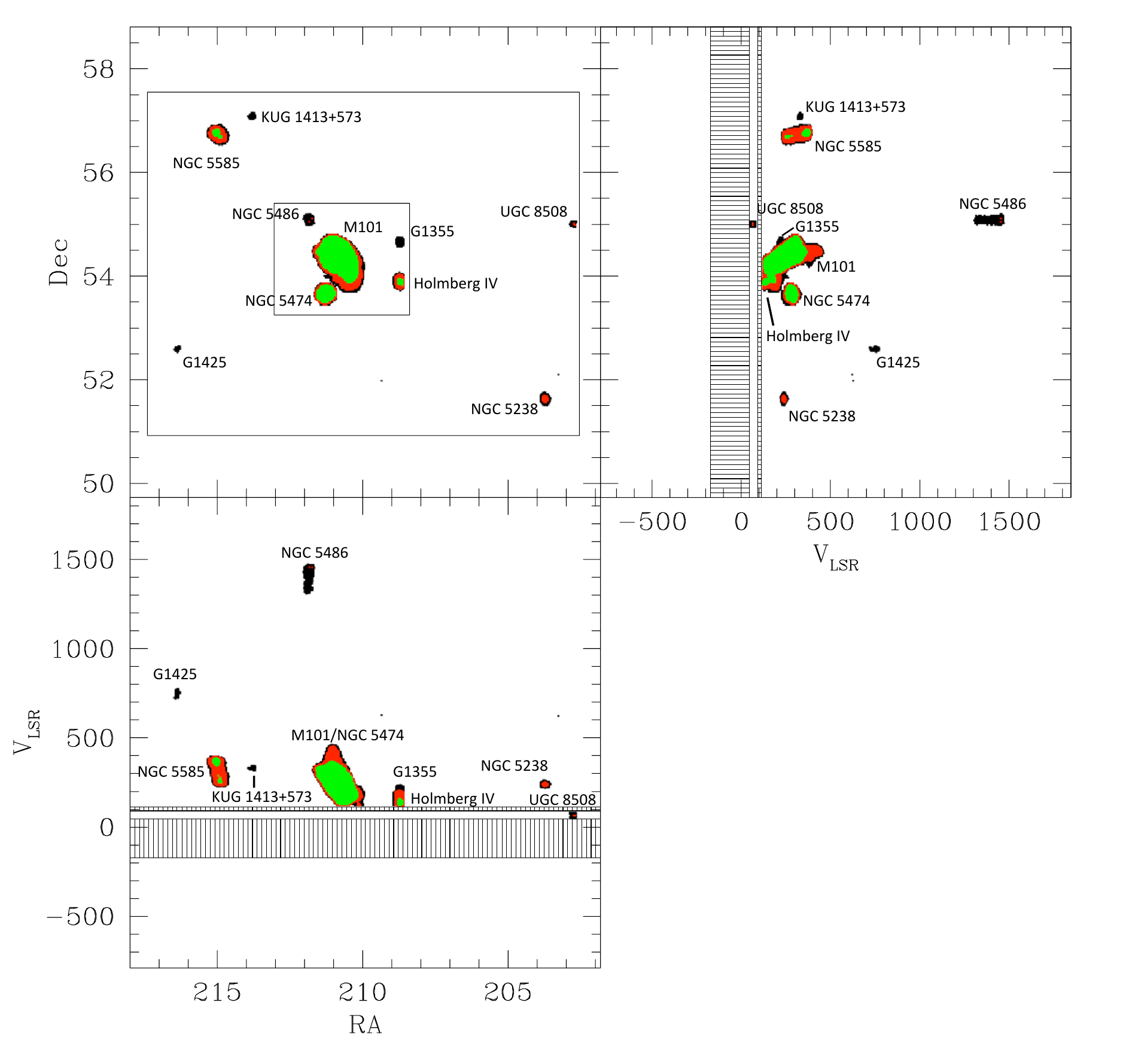}} 
\caption{The distribution of detected HI in the M101 field. {\it Upper
left:} the location of significant HI detections on the sky. The outer
rectangle shows our total survey field, while the inner rectangle shows
the inner field surveyed at higher sensitivity. Green shows HI with peak
emission at the 100$\sigma$ level, red shows the 10$\sigma$ level, and
black shows HI at the 3.5$\sigma$ level. {\it Upper right:} the
\vlsr-declination projection of detected HI. The hashed regions show
regions contaminated by narrowband RFI (\vlsr$=91.4-117.1$ \kms) and
Galactic emission (\vlsr$=-171.4 - +45.0$ \kms) where HI detection was
not done. {\it Lower left:} the RA-\vlsr\ projection of the HI
distribution.}
\label{environ}
\end{figure*}

\begin{figure}[]
\centerline{\includegraphics[width=3.5truein]{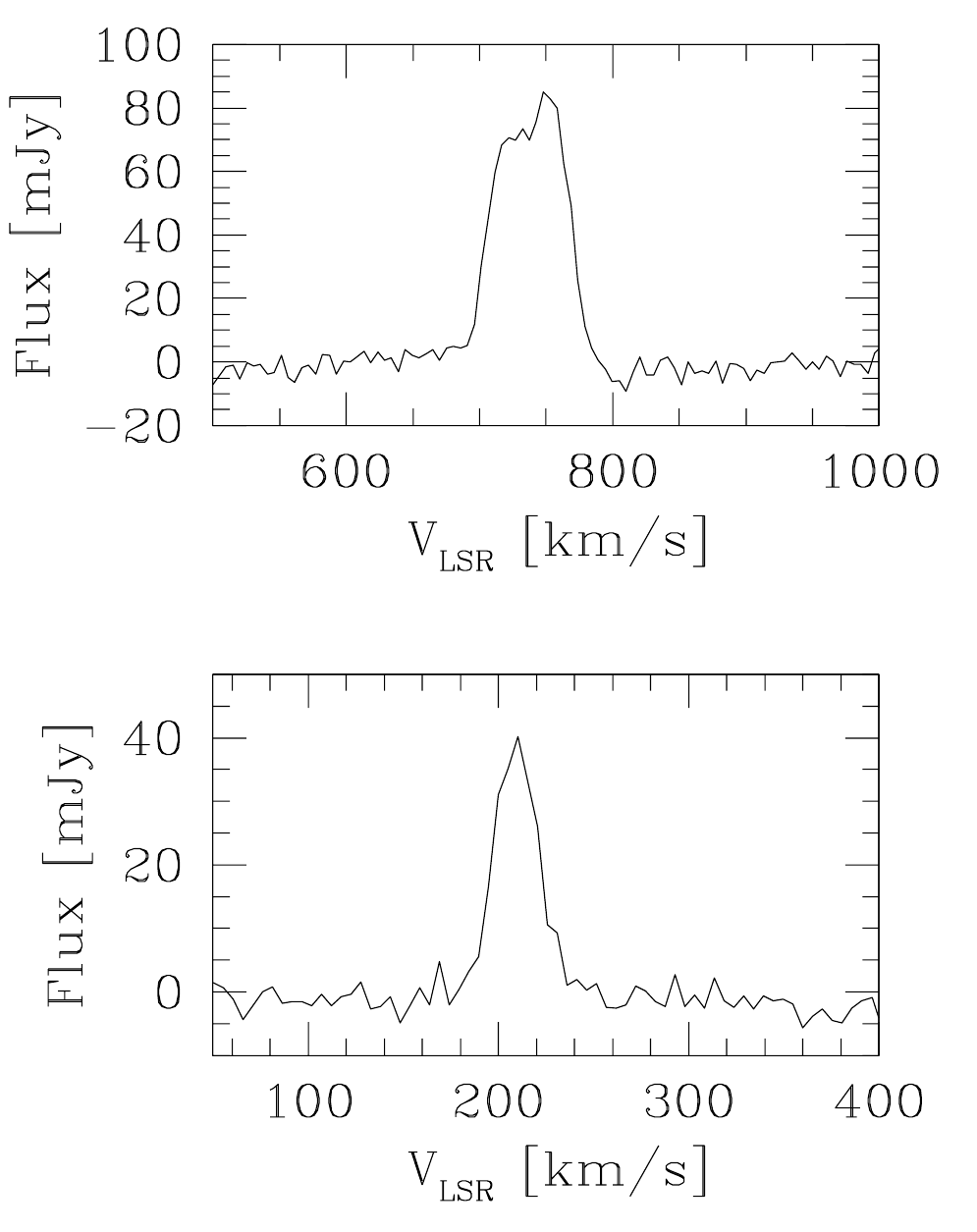}}
\caption{Spectra of the new HI detections. Top:
GBT 1425+5235. Bottom: GBT 1355+5439.}
\label{newspectra}
\end{figure}

The distribution of detected HI emission is shown in
Figure~\ref{environ}. We detect all known galaxies in the survey field
with velocities in our detection range, along with two new sources,
dubbed GBT 1425+5235 (hereafter G1425) and GBT 1355+5439 (hereafter
G1355).\footnote{While we use IAU-recommended nomenclature for the
source names in this paper, the authors prefer the names ``Skipper" (G1425)
and ``Gilligan" (G1355) for these objects, adopted in honor of Sherwood
Schwartz, who passed away during the writing of this paper.} To verify
these sources, we conducted short, follow-up position-switched
observations of each source using the GBT spectrometer in 2012 September.
In $\sim$ 30 minute integrations, we achieved an rms sensitivity
of $\sim$ 3 mJy beam$^{-1}$ and easily recovered both objects. Our HI
spectra of these sources from the follow-up observations are shown in
Figure~\ref{newspectra}, and their integrated properties are given in
Table~\ref{newprops}. Both of these sources are unresolved in our data
due to the large 9.1\arcmin\ (18 kpc) beam of the GBT.

\begin{figure*}[]
\centerline{\includegraphics[width=5.0truein]{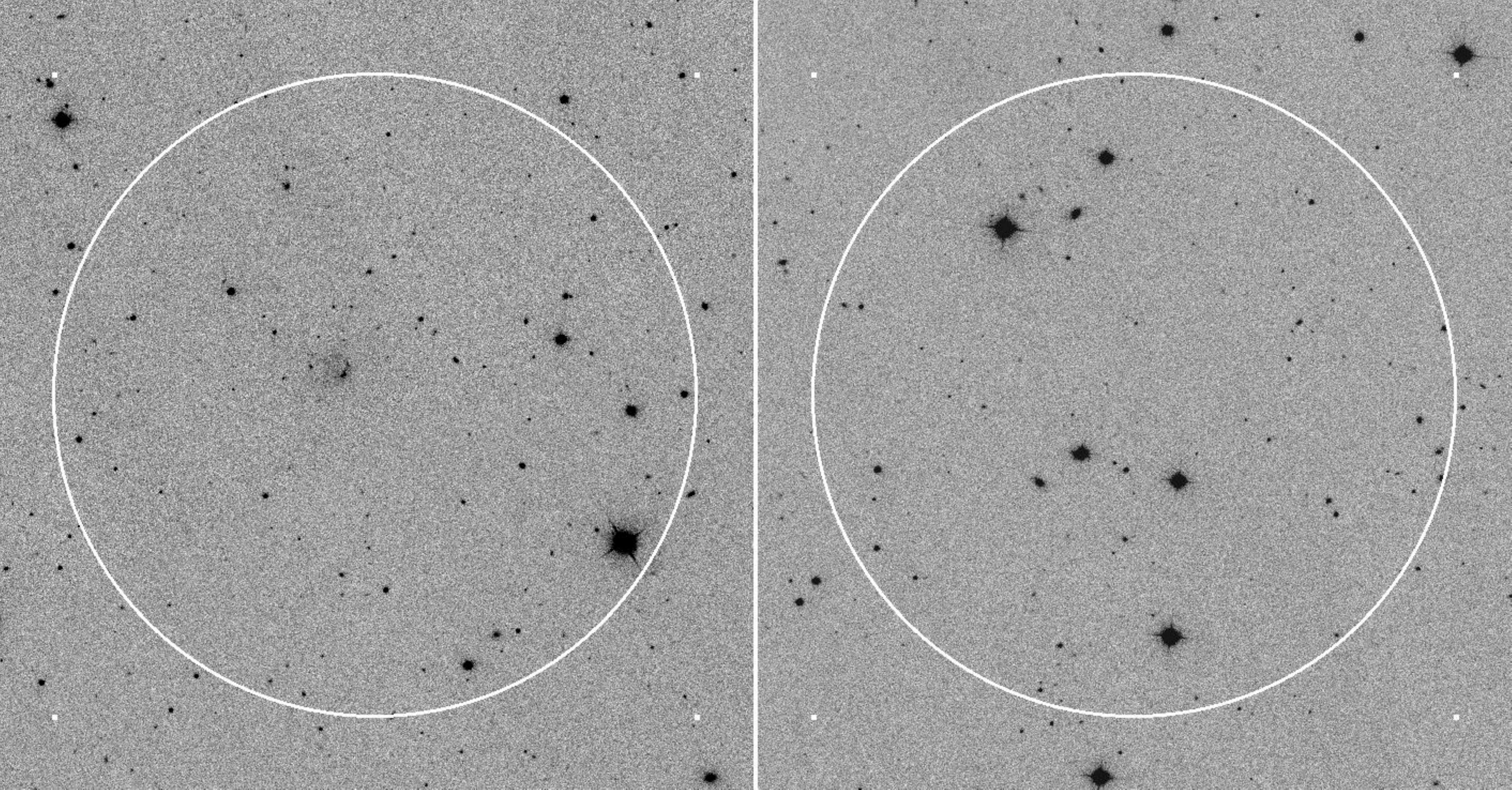}}
\caption{SDSS $g$-band images at the positions of the new HI detections.
Left: GBT 1425+5235 (G1425). Right: GBT 1355+5439 (G1355). The limiting
surface brightness of the SDSS images is $\mu_{g,{\rm lim}} = 26.5$, and
the white circles indicate the 9\arcmin\ GBT beam size. G1425 shows a
diffuse, low surface brightness optical counterpart near the center of
the GBT beam. No plausible counterpart is visible for G1355; the objects
within the GBT beam are either stellar or background galaxies at $z \sim
0.1$.
}
\label{newimages}
\end{figure*}

G1425 has a cataloged optical counterpart WHI J1425+52, identified by
Whiting \etal (2007) as Galactic nebulosity based on relatively shallow
R-band imaging with a limiting surface brightness of $\sim$ 25.5.
However, with a significant HI detection, a velocity of \vlsr=736 \kms,
and a line width of \W20=77 \kms, G1425 is unlikely to be Galactic
in origin. To examine G1425's optical properties in more detail, we
used deeper ($\mu_{g,lim} \sim 26.5$) Sloan Digital Sky Survey Data Release 7 
(SDSS DR7) imaging (Abazajian \etal
2009) of the field, shown in Figure~\ref{newimages}, to characterize the
morphology and colors of the optical counterpart. The system consists of
a few high surface brightness knots asymmetrically embedded in a
diffuse, low surface brightness ($\mu_g = 24.8$) structure approximately
0.65\arcmin\ in diameter. Within this aperture, the object has a total
magnitude of $g=17.24$; if at the adopted distance of M101, it would
have an absolute magnitude of M$_g = -12.0$ and a diameter of 1.3 kpc. Its
very blue colors (see Table \ref{newprops}) and knotty appearance are
characteristic of a star forming dwarf irregular galaxies (\eg van Zee
2001). The closest Local Group analogue to G1425 might be the Pegasus
dwarf irregular galaxy (Gallagher \etal 1998; Kniazev \etal 2009),
although Pegasus is more symmetric, higher in central surface
brightness, and redder than G1425. In these properties, G1425 is
similar to the extreme low surface brightness dwarf galaxies (Bothun \etal 1998;
van Zee 2001).

In contrast to G1425, around G1355 we find no plausible optical
counterpart down to the $\mu_{g, lim} \sim 26.5$ limiting surface
brightness of the SDSS imaging (shown in Figure~\ref{newimages}). There
are no extended sources present in the field, and the two small galaxies
within the 9.1\arcmin\ GBT beam are background galaxies (SDSS
J135459.84+543736.8 at $z=0.132$ and SDSS J135456.31+544122.6 at
$z=0.075$). The source also lies precisely on the edge of the field in
our deep imaging of M101 (Mihos \etal 2012), which has an even deeper
limiting surface brightness of $\mu_B = 29.5$, but again there is no
detection.

The lack of any optical detection around G1355 makes classifying this
source difficult. One possibility is that G1355 is a high velocity
cloud (HVC) in the Galactic halo. The M101 field lies near (but not
within) the Galactic HVC complex C (Wakker \& van Woerden 1991), but
this complex has velocities in the range of \vlsr= --235 to --50
\kms, very dissimilar to G1355's velocity of \vlsr=+210 \kms. Also,
in our data G1355 is also not part of any larger HI structure
identifiable with Galactic emission --- it is unresolved in the GBT beam
with no surrounding diffuse emission, and the nearest velocity channel
with {\it any} diffuse emission comes in at \vlsr=+45 \kms. Furthermore,
no HVCs have been identified in this part of the sky with velocities
greater than +50 \kms (Wakker \& van Woerden 1991).

If instead of being a Galactic HVC, G1355 is at the M101 distance, it
would have a mass of $10^7$ \Msun\llap, and with a velocity width of
\W20=41 \kms\ it would be similar to the HI clouds found around
the interacting galaxies M81/M82/NGC3077 (Chynoweth \etal 2008).
However, those clouds are found embedded in the extensive HI tidal
debris surrounding the galaxies (Yun \etal 1994) and have likely formed
from HI stripped out of the galaxies due to the group interaction.
Although G1355 could also have formed through an interaction between
M101 and one of its companion galaxies, perhaps Holmberg IV, we see no
diffuse HI structures connecting G1355 to either M101 (projected 80
kpc to the SSE) or Holmberg IV (50 kpc to the south). 

Aside from G1425 and G1355, we find no other isolated HI sources
aside from the well-known galaxies in the field. As both G1425 and
G1355 have masses 5-10x higher than our 90\% detection threshold in
the outer field, other clouds of similar mass should have been easily
detected as well. Other than G1355, we find no evidence for any
robust population of massive HI clouds in the M101 group, similar to the
findings of other searches for starless HI clouds in the group
environment (de Blok \etal 2002; Pisano \etal 2007, 2011; Kova\v c \etal
2009; Chynoweth \etal 2009). Any undetected population of discrete HI
clouds in the M101 group must have masses less than our detection limit
of $\sim 10^6$ \Msun.

\begin{deluxetable}{lcc}
\tabletypesize{\scriptsize}
\tablewidth{0pt}
\tablecaption{New HI Detections}
\tablehead{\colhead{ } & \colhead{GBT 1425+5235} & \colhead{GBT 1355+5439}}
\startdata
RA & 14:25:29.3 & 13:54:50.6 \\
Dec & +52:34:59 & +54:38:50 \\
D$_{\rm M101}^a$ [kpc] & 450 & 160 \\
$v_{sys}$ [\kms] & 737 & 210 \\
\W20 [\kms] & 77  & 41 \\
HI flux [Jy \kms] & 5.1 & 1.1 \\
M$_{\rm HI}^b$ [\Msun] & $5.7\times10^7$ & $1.2\times10^7$ \\
$g$ & 17.24 & ... \\
$u-g$ & 0.30 & ... \\
$g-r$ & 0.30 & ... \\
$r-i$ & 0.07 & ... \\
M$_g$ & $-$11.95 & ... \\
L$_{\rm B}^c$ [\Lsun] & $7.4\times 10^6$ & ... \\
M$_{\rm HI}$/L$_{\rm B}$ & 0.25 & ... 
\enddata
\tablecomments{a) D$_{\rm M101}$ is the projected physical separation
between the object and M101. b) M$_{\rm HI}$ calculated assuming the
objects are at the adopted M101 distance of 6.9 Mpc. c) L$_{\rm B}$
calculated using the conversion from $ugriz$ to B from Lupton (2005).}
\label{newprops}
\end{deluxetable}

\subsection{Extended HI around M101}

A variety of studies have focused on the HI properties of M101,
including both single-dish observations (HW79) and array synthesis
observations (Allen \& Goss 1979; vdHS88; Walter \etal 2008). Rather
than focus our discussion on the well-studied inner regions of M101, we
will concentrate on studying the diffuse outer features of the HI
distribution; it is here that our very deep, wide-field data adds new
information to the M101 system.

\begin{figure*}[]
\centerline{\includegraphics[width=6.5truein]{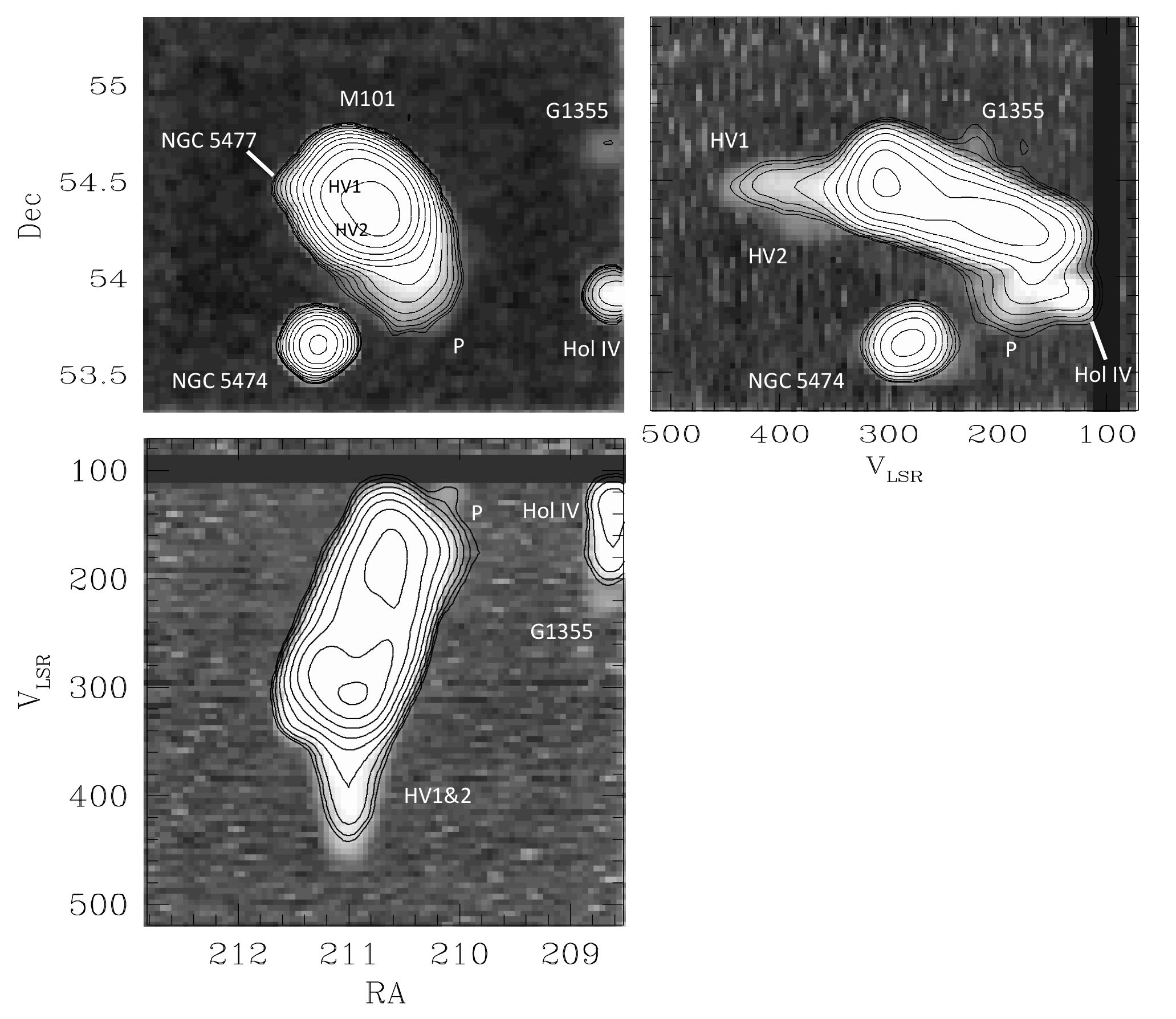}}
\caption{
Projected views of the inner, high sensitivity portion of the M101
datacube. To best highlight diffuse features, we show the peak flux
(rather than integrated flux) along each projected axis. Top left: HI
intensity map; contours of column density are shown which run from
$\log ({\rm N_{HI}})=$18 to 20.5 in steps of 0.25. Top right: Declination vs \vlsr. Bottom left: \vlsr\ vs right
ascension. Contours in the position-velocity diagrams running logarithmically
from 0.1 to 50  Jy beam$^{-1}$, and the bright bands near \vlsr=100 
are channels corrupted by narrowband RFI. Galaxies are
labeled; ``HV1" and ``HV2" refer to high velocity clouds first
identified by van der Hulst \& Sancisi 1988, and ``P'' refers to the
southwest HI plume.
}
\label{cubepv}
\end{figure*}

\begin{figure*}[]
\centerline{\includegraphics[width=7.0truein]{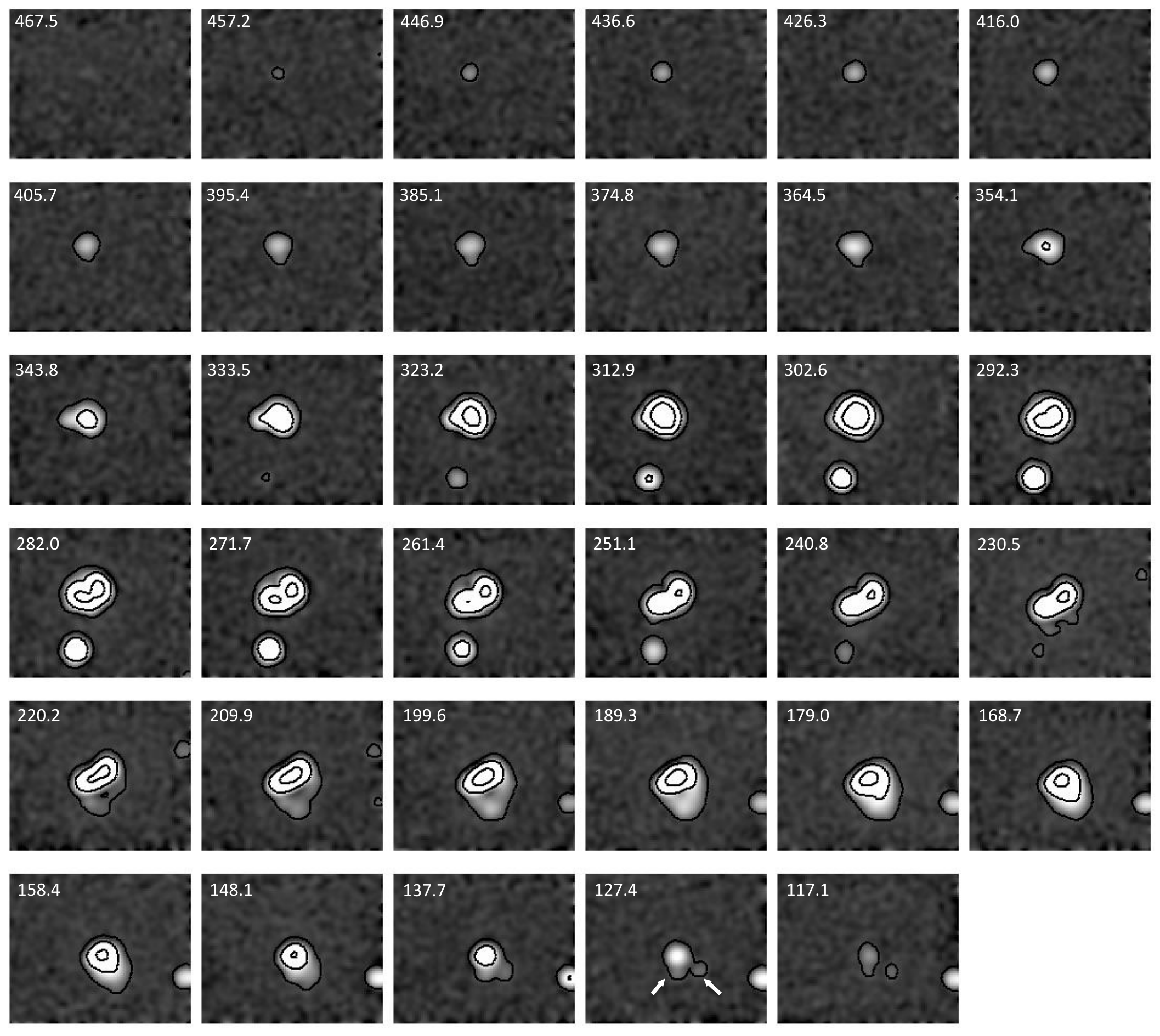}}
\caption{
Channel maps of the inner, high sensitivity portion of the M101
datacube. Channel velocities are given in the upper left corner of each
panel; and contours are shown at 0.001, 0.05, 0.01, 0.5, 1.0, and 2.0 
Jy beam$^{-1}$. The faintest features visible in the channel maps have an
intensity of about 0.1 Jy \kms\ or $\log ({\rm N_{HI}})=17.0$. The
extended southwest loop of HI can be seen in channels with
\vlsr=220--158 \kms, and the two low mass HI clouds found near the
southwest loop are marked at \vlsr=127.4 \kms. G1355 can also be seen
very faintly on the upper right edge of the frame near \vlsr=220 \kms.
}
\label{channels}
\end{figure*}

Figure \ref{cubepv} shows three orthogonal projections of our full HI
datacube. In this figure, the intensity greyscale has been stretched to
best show the diffuse HI structures at large radius and low column
density, which saturates much of the emission from the inner parts of
the galaxy. In our deep HI intensity map, the most striking feature is
the long plume of HI extending to the southwest of M101. This gives the
galaxy an HI asymmetry at low column density which is in the opposite
sense from that indicated by the high column density THINGS map and at
low surface brightness in the optical (Mihos \etal 2012), both of which
show M101 extended to the northeast.\footnote{At even higher column
density in the inner few kpc of the galaxy, the HI asymmetry swings back
{\it again} to the southwest side of the galaxy, mirroring the
well-known lopsidedness of M101's optical disk (see, \eg Mihos \etal
2012).} This feature was noticed at higher column density ($\sim 4\times
10^{18}$ cm$^{-2}$) by HW79, and the HI synthesis map of vdHS88 shows it
connecting smoothly into M101's disk at yet higher column density ($1.4
\times 10^{19}$ cm$^{-2}$). In our data, we trace this plume even
farther out in radius to $\sim$~85 kpc from the center of M101, where
the column density has dropped to $\sim 3\times 10^{17}$ cm$^{-2}$. The
plume has no optical starlight associated with it, down to a limiting
surface brightness of \mub=29.5 (Mihos \etal 2012). The
position-velocity plots in Figure \ref{cubepv} and the channel maps in
Figure \ref{channels} show that this plume co-rotates with M101,
following the general decline in M101's rotation curve on the southwest
side of the disk. However, the plume doesn't show the simple kinematic
behavior expected for simple circular motion; instead, it takes on a
loop-like appearance in the channel maps from \vlsr=210 to 180 \kms,
before filling in with emission at lower velocity of \vlsr=150 \kms.
This could be a signature of a warping or flaring of the plume out of
the plane of rotation, and the complex kinematic structure of this
feature leads to the diffuse, irregular HI emission seen around the main
kinematic structure of the plume in the position-velocity maps in Figure
\ref{cubepv}.

In addition to the plume itself, we see a number of discrete HI features
in the dataset. Near the southwest plume, but discrete in velocity space
from the plume itself, we find two distinct HI clouds. These clouds show
up in the last few channel maps shown in Figure \ref{channels} around
\vlsr=127 \kms\ and have HI masses $\sim 10^7$ \Msun\llap. An
examination of the extremely deep optical imaging by Mihos \etal 2012
shows no resolved optical counterparts at the position of these clouds.
Given their low HI masses, it is unlikely that they could have pulled
such a long plume of HI from M101 itself, but they may be similar to the
HI clouds seen in the tidal debris in the M81 group, thought to form
from tidally stripped gas during strong galaxy interactions (Chynoweth
\etal 2008, 2011).

The position-velocity projections of the datacube (Figure \ref{cubepv})
also show other HI features kinematically distinct from the ordered
disk rotation of M101. The two regions of high velocity gas
first noted by vdHS88 can easily be seen as a bright finger of HI
emission extending to \vlsr=460 \kms\ and another fainter feature just
to the south at \vlsr=400 \kms. Both these features are projected onto
the outskirt's of M101's disk and are spatially coincident with each of
the low surface brightness tidal features seen in the outer optical disk
(Mihos \etal 2012). Several dynamical models have been proposed to
explain this high velocity gas, including infall of gas clouds onto M101's
disk (vdHS88) and tidal perturbations due to the interaction with
M101's dwarf companion galaxy NGC 5477 (Combes 1991).

\begin{figure*}[]
\centerline{\includegraphics[width=6.5truein]{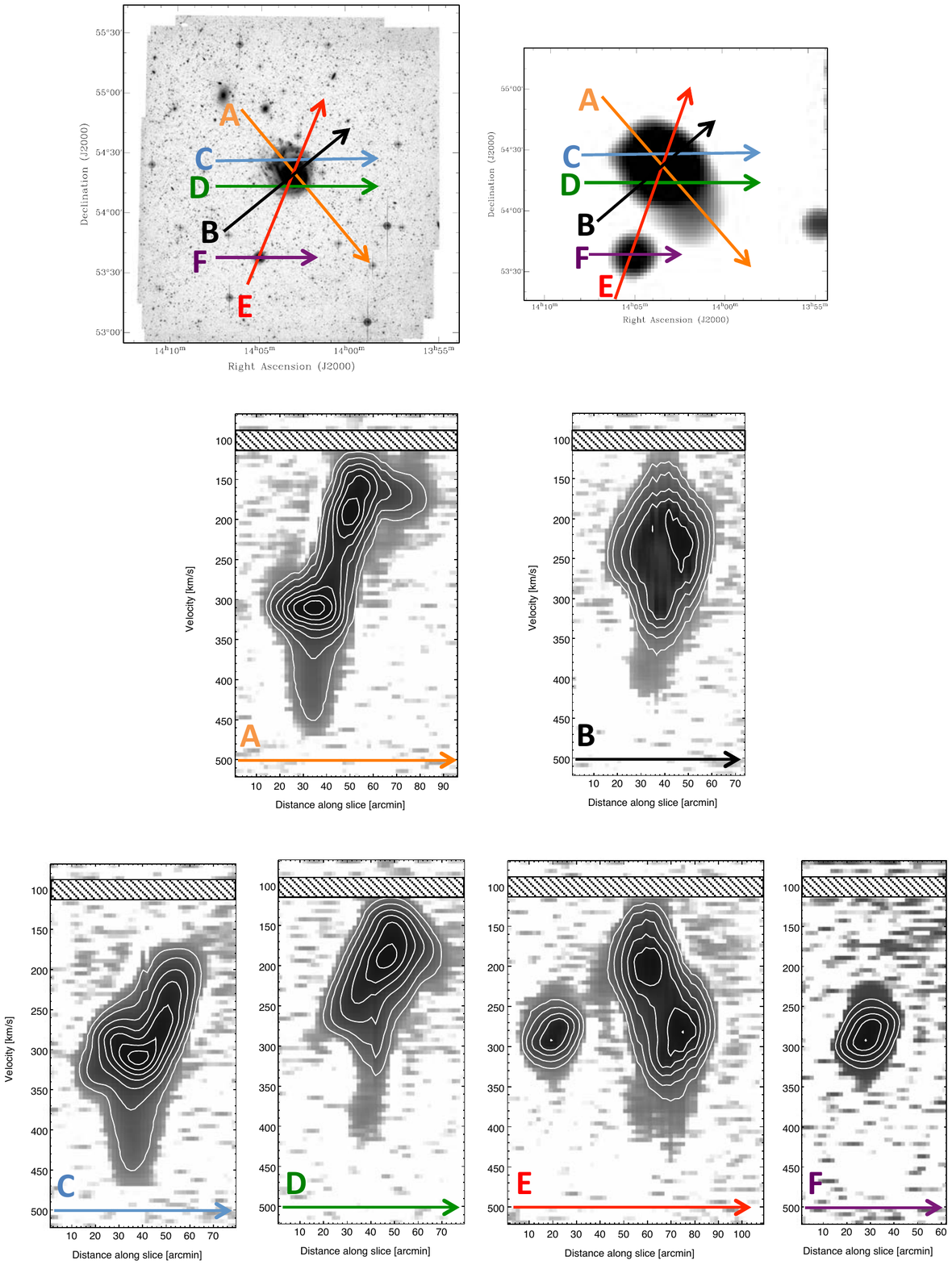}}
\caption{
Position-velocity slices through M101. The top panels show a wide field
B-band image of M101 from Mihos \etal (2012) along with our GBT HI
intensity map, with four spatial slices defined. The middle panels show
slices through the kinematic major (A) and minor (B) axes of M101
($\psi_{kin}$=39\degr; Bosma \etal 1981). The lower panels show the
position-velocity slice along four other spatial cuts. Slice C is a
slice at constant declination through NGC 5477 and the northern part of
M101's disk. Slice D is a slice at constant declination through the
southern part of M101's disk. Slice E is a slice along the line
connecting NGC 5474 and M101. Slice F is a slice along the major axis of
NGC 5474. The faintest structures visible in the slices have an
intensity of approximately 3 mJy beam$^{-1}$ ($\sim 1\sigma$), while the
greyscale saturates to black at $\sim$ 1 Jy beam$^{-1}$. Contours are
shown at 0.05, 0.2, 0.6, 1.3, 2.3, 3.5, and 5 Jy beam$^{-1}$, and the
hatched areas are velocity channels blanked due to narrowband RFI.
}
\label{pvplots}
\end{figure*}

To further explore the HI kinematics around M101, Figure \ref{pvplots}
shows position-velocity slices along selected spatial cuts through the
data cube. The slices are extracted using the {\tt kpvslice} task in the
Karma data analysis package (Gooch 1996) and are 1.65\arcmin\ (one
spatial pixel) wide. Slice A is a cut along the kinematic major axis
($\psi_{kin}=39\degr$, Bosma \etal 1981), showing M101's projected HI
rotation curve. This cut runs through the high velocity gas on the
northeast side of the disk as well as the extended HI plume to the
southwest. On both sides of the galaxy the projected velocities show a
significant decline at larger radius, which could be a signature of
either a declining rotation curve or a warping of the disk. While M101's
rotation curve does show a modest drop in the rotation speed in the
inner disk ($r=$10\arcmin--14\arcmin; Bosma \etal 1981), we are seeing
this behavior at much larger radius ($r \sim$ 30\arcmin). A minor axis
cut can help discriminate between a warp and an overall decline in the
rotation curve. In pure circular rotation, a minor axis cut should show
no gradient, while a warp would skew the kinematic axes with radius and
result in a gradient along a fixed kinematic minor axis cut. We show a
minor axis cut in slice B, which does reveal a very modest gradient, but
given the beam smearing in our data and the presence of the high
velocity gas in the cut it is hard to interpret this gradient
unambiguously. However, we also note that the projected velocities along
the major axis show are asymmetric -- they drop more steeply on the
southwest side of the disk, and along this portion of the cut, diffuse
gas can be seen moving both faster and slower than the main spine of the
rotation curve. This is the region of the extended HI plume and again is
suggestive of diffuse gas moving out of the disk plane, perhaps in a
warp.

Aside from the major and minor axis cuts, we also present a variety of spatial
cuts aimed at exploring the kinematics of the M101 system on larger scales.
Slices C and D are cuts at constant declination through M101, and are analogous to
the cuts in the HI synthesis map shown in Figure 2 of vdHS88; slice C
cuts through the dwarf companion NGC 5477 and then through the north
side of the disk, while slice D cuts through the south side of the
disk. In slice C, NGC 5477 can be seen at the beginning of the cut at
\vlsr=330 \kms, redshifted 50 \kms\ with respect to M101.

Slice E cuts first through the nuclei of both NGC 5474 and M101, and
shows the presence of diffuse HI gas located between the galaxies. This
gas is at velocities intermediate between NGC 5474's systemic velocity
and the velocity of the southeastern edge of M101's disk, suggestive of
material drawn out of M101 due to an interaction with NGC 5474. Indeed,
a cut along the kinematic major axis of NGC 5474 (slice F) shows a
slight kinematic asymmetry as well; there is more gas on the lower
velocity side of the profile, at similar velocity (\vlsr=240 \kms) to
the diffuse gas found between the galaxies in slice E. The gas at
intermediate velocity may represent an actual bridge of material between
the galaxies, or simply be a plume of tidally stripped gas that lies
between the galaxies in projection.

\section{Discussion}

Our deep HI imaging of the M101 group does not reveal any large
population of HI clouds at masses of a few $\times 10^6 - 10^7$ \Msun,
echoing the results from HI searches in other nearby galaxy groups (de
Blok \etal 2002; Pisano \etal 2007, 2011; Kova\v c \etal 2009; Chynoweth
\etal 2009). Of the two new sources we detect in the group, one (G1425)
is optically identifiable using SDSS imaging as an extremely low surface
brightness dwarf galaxy. G1425's properties make it comparable to the
gas-rich HI dwarfs being found in blind HI surveys such as ALFALFA
(Haynes \etal 2011; Cannon \etal 2011; Huang \etal 2012), with large HI
fractions, blue colors, irregular structure, and evidence for on-going
star formation. In contrast, G1355 potentially represents a much rarer
object --- a starless HI cloud with mass \MHI$=10^7$ \Msun. While such
objects are seen in the tidal debris surrounding strongly interacting
galaxies (\eg Chynoweth \etal 2009), G1355 itself seems to be truly
distinct and isolated from M101's extended (and likely tidal) HI plume.
We detect no extended HI emission around G1355 out to a radius of
35\arcmin\ (70 kpc); given that G1355's peak brightness (40 mJy
beam$^{-1}$) is an order of magnitude greater than the background noise,
this places a lower limit on the HI column density contrast between
G1355 and its environment of $\sim$ 10:1. In comparison, the HI clouds
we find near M101's tidal plume, while comparable in mass to G1355, are
embedded in the more diffuse plume emission with a density contrast of
$\sim$ 3:1, and are very likely tidal in origin. Isolated starless HI
clouds like G1355 are extremely rare --- for example, less than 2\% of
the HI detected objects in the ALFALFA survey have no associated optical
counterpart in SDSS or DSS2 imaging (Haynes \etal 2011). Many (3/4) of
the starless ALFALFA HI sources are found in fields with known objects
(Haynes \etal 2011) and may be associated with tidal debris or extended
HI. While G1355 is also found in a field with neighboring galaxies, in
this case we see no evidence tying it to any of the extended HI in the
M101 group.

Given that G1425 and G1355 are found in the loose group environment,
there may be a significant degree of photoionization from the UV
radiation emitted by M101, as well as photoevaporation from any
intragroup medium (e.g. Benson \etal 2002).\footnote{While the M101
group does not have a dense X-ray emitting IGM like those seen in
elliptical-dominated groups, a more tenuous hot halo capable of
stripping low mass dwarfs remains a possibility (\eg Lin \& Faber 1983;
Freeland \& Wilcots 2011).} Because of this heating, it is expected that
such low mass dwarfs may be surrounded by an envelope of ionized gas
that shields the HI gas from further disruption (Sternberg \etal 2002).
The detectable HI, then, may be just the tip of the iceberg that marks
the true gas content associated with these objects. Since these dwarfs
will also be subject to ram pressure stripping by the intragroup medium,
it may well be true that they are newly accreted into the M101 group, or
on more circular orbits that prevent them from encountering particularly
dense intragroup gas (e.g. Gunn \& Gott 1972; Nichols \& Bland-Hawthorn
2011; Freeland \& Wilcots 2011).

More speculatively, it might be tempting to call G1355 a
pre-reionization fossil dwarf, a protogalaxy whose star formation
history was cut short by the epoch of reionization (Ricotti \& Gnedin
2005). While the canonical fossil dwarf is presumed to be an ancient
dSph that is nearly devoid of gas, more detailed work predicts that
there is a class of 'polluted' fossil dwarfs that have a late spurt of
gas accretion from the intergalactic medium onto an essentially dark minihalo (Ricotti
2009). These pre-reionization artifacts, then, should be undetectable
optically, but should show up in a deep HI survey due to their cold gas
content (Bovill \& Ricotti 2011) --- similar on the face to G1355.
However, the likeliest location of polluted fossil dwarfs is in voids,
not loose groups. In addition, with a line width of \W20=40\kms,
G1355 might exceed the minimum halo mass below which reionization
could have halted collapse and star formation in the dwarf (M$_{min}
\sim 10^8 - 10^9$\Msun\ ($v_{\rm circ}=20$ \kms) (Babul \& Rees 1992;
Efstathiou 1992) and thus not be a good candidate for a pre-reionization
fossil dwarf.

If both G1425 and G1355 are virialized, HI-dominated dwarf
galaxies, we can use their HI mass to place them on the baryonic
Tully-Fisher relationship (McGaugh \etal 2000; McGaugh 2012). There is
significant scatter at the low mass end of the relation, for G1425 we
get \W20=30--75 \kms\ and for G1355 \W20=20--50 \kms. Our
measurements of \W20 (77 \kms\ and 41 \kms, respectively) are consistent
with, but on the high side of, these ranges. This may argue that both
systems are observed largely edge on, which maximizes the rotation
width, or that the objects have significant non-circular motion, either
because they are not yet virialized, or because they are being tidally disrupted
by M101 and other galaxies in the M101 group. Followup HI synthesis
mapping would help discriminate between these possibilities and shed
light on the true nature of these objects.

The lack of a large population of neutral hydrogen clouds in the M101
group argues that M101 is likely not experiencing significant cold accretion
directly from the surrounding intragroup medium. This is perhaps not
surprising for a massive galaxy in the local universe. M101's
circular velocity ($\sim$ 220 \kms; Bosma \etal 1981) and size
suggest a total mass on the order of $10^{12}$ \Msun\llap. This mass
marks the threshold above which the galaxy accretes through a
predominantly hot, shock-heated mode, where cold flows are unlikely to
reach the disk unimpeded by the hot halo. However, Kere\v s \etal 2009
find that HI filaments and clouds can still form via a density inversion as
the the accreting gas enters a hot gaseous halo, triggering Rayleigh-Taylor
instabilities. These clouds rain down radially on the disk from
distances of $\sim 50$ kpc and the most massive of these $\sim 10^6$
\Msun are expected to survive and feed the disk (Kere\v s \etal 2009).
With N$_{\rm HI}$ $>$ 10$^{16}$ cm$^{-2}$, a covering fraction approaching
10\%, and velocities distinctly decoupled from disk rotation, this
phenomenon could be a channel to form HVCs (Stewart \etal 2011). 

While this hot accretion mechanism can continue feeding gas to the disk
of a massive galaxy like M101, such clouds would be below our detection
threshold and go unnoticed in our dataset. Indeed, the masses of
G1355 and the two discrete clouds seen near the extended HI plume are
all $\sim 10^7$ \Msun, higher than expected for clouds condensing
out of a hot halo. At least for the two clouds near the plume, it is
more likely that they formed through the same tidal interaction that led to
the plume itself, much like the HI clouds found in the M81/M82 system
(Chynoweth \etal 2008). While G1355 may have formed this way as well, 
the lack of a connection to any of the diffuse HI tidal debris around
M101 makes this identification less secure.

While the detected cloud masses in M101's plume are not a good match for
clouds accreting from a warm halo, they are similar to the masses of the
high velocity gas complexes seen in M101's outer disk ($10^7-10^8$
\Msun\llap, vdHS88). A large fraction of the material tidally stripped
during galaxy encounters can fall back to the host galaxy over long
timescales (Hibbard \& Mihos 1995); were these clouds to fall back on to
M101, they too could punch through the disk gas and lead to the
formation of such high velocity cloud complexes; mass loading during the
interaction between the cloud and the disk gas could increase the mass
of the resulting high velocity gas complex as well. Thus the clouds we
see may well be the progenitors to the current high velocity gas
features seen in M101's disk, spawned by a past interaction that also
led to M101's strong asymmetry and extended HI plume.

M101's well-known asymmetry has long been believed to arise from an
interaction (Beale \& Davies 1969; Rownd \etal 1994; Waller \etal 1997),
but unambiguously identifying the interacting partner has remained
difficult. The two most obvious candidates are NGC 5477, located
22\arcmin (44 kpc) east of M101, and NGC 5474, 44$\arcmin$ (88 kpc) to
the south. NGC 5477 resides near M101's crooked NE spiral arm; while it
is clearly distorted and interacting with M101, its low luminosity
(M$_V =-15.3$; de Vaucouleurs \etal 1991) makes it unlikely to be massive
enough to produce such a strong morphological response in a giant Sc
spiral like M101. NGC 5474 is a more luminous system (M$_V =-18.4$; de
Vaucouleurs \etal 1991) and sports a strongly asymmetric disk (Rownd
\etal 1994, Kornreich \etal 1998), but shows no sign of
any extended tidal debris in the optical (Mihos \etal 2012). Previous
single dish mapping of the M101 system hinted at the possibility of an
HI bridge between NGC 5474 and M101 (HW79), but
subsequent HI synthesis imaging has shown no sign of any HI tidal
features around NGC 5474 (Rownd \etal 1994, Kornreich \etal 2000) which
might trace a previous interaction. 
Our data strengthen the case for a past interaction between M101 and NGC
5474. In diffuse HI, we find gas between the galaxies at intermediate
velocity, as would be expected from a tidal feature drawn out of M101 by
the interaction. First hinted at in the single dish data of HW77, this
feature can clearly be seen in slice D on our Figure~\ref{pvplots}. NGC
5474 itself shows a slight kinematic asymmetry as well (slice E), with
more extended HI emission at lower velocities similar to that on the
southern edge of M101's disk.

An accurate estimate of the mass ratio of the M101/NGC5474 pair is quite
difficult. Because both galaxies are largely face-on, and because of the
kinematic irregularity of M101's extended disk, deriving dynamical
masses for each galaxy is fraught with uncertainty (HW77; Allen \etal 1978;
Bosma \etal 1981; Rownd \etal 1994), and estimates of the mass ratio
between the pair range from 20:1 to 200:1 depending on the different
dynamical mass estimates used. Even under the simplest (and likely
incorrect) assumption that the galaxies have similar mass to light
ratios, the mass ratio of the pair can range from 15:1 based on the
V-band flux ratio (RC3, de Vaucouleurs \etal 1991) to 40:1 based on the
K-band flux ratio (Jarrett \etal 2003). Even with the significant
uncertainty, and even accounting for any mass loss that may have been
experienced by NGC 5474 during the encounter, it seems clear that this
would have been a high mass ratio interaction. For this type of encounter
to do such significant damage to M101's disk, the encounter would have 
to be largely prograde, or involve a close impact parameter, since distant or
retrograde encounters typically result in a relatively weak disk response (Toomre
\& Toomre 1972). Close or prograde encounters would also be effective at drawing
material out of the galaxy disks and forming the kinematic
irregularities, extended HI plume, and high velocity cloud structures we
see in our data.

\section{Summary}

Taken as a whole, our deep, wide-field HI map of the M101 group has
revealed a number of features indicative of a very dynamic gaseous
environment surrounding M101. We have traced M101's extended (and likely
tidal) plume of neutral hydrogen 100 kpc to the southwest of the galaxy,
where it contains two distinct, massive ($\sim 10^7$ \Msun) clouds. Clouds
such as these may be the source of gaseous infall that gives rise to the
high velocity gas seen in M101's outer disk (vdHS88), as they fall back
onto the galaxy from the tidal debris. We also identify two new discrete
HI sources in the M101 group, one a new and extremely low surface brightness dwarf galaxy
and another a starless HI cloud or, possibly, primordial dwarf galaxy.
Finally, we identify a possible bridge or plume of gas at intermediate
velocities between M101 and NGC 5474 which may indicate a recent
interaction between the galaxies, possibly leading to their highly
asymmetric morphologies and giving rise to M101's extended HI plume.

With growing evidence that both NGC 5477 and NGC 5474 have recently
interacted with M101, we are building a picture of the M101 group as a
dynamically active system. If our two new HI detections do prove to be
gas-rich, relatively unevolved dwarf galaxies, the fragility of these
objects in the group environment may further underline the image of a
still assembling galaxy group.

\acknowledgments

J.C.M.'s work on this project has been funded by the NSF through grants
AST-0607526 and AST-1108964. K.M.K. acknowledges the NRC Research
Associateship program for funding support. K.H.B. acknowledges support
from NSF CAREER award AST-0807873, and thanks the Aspen Center for
Physics (supported through NSF grant 1066293) for hospitality and a
lovely office to puzzle over p-v diagrams. D.J.P. acknowledges support
from NSF CAREER grant AST-1149491. Basic research in radio astronomy at
the Naval Research Laboratory is supported by 6.1 base funding. We also
thank Stacy McGaugh for several helpful discussions.

This research has made use of the NASA/IPAC Extragalactic Database (NED)
which is operated by the Jet Propulsion Laboratory, California Institute
of Technology, under contract with the National Aeronautics and Space
Administration. Funding for the SDSS and SDSS-II has been provided by
the Alfred P. Sloan Foundation, the Participating Institutions, the
National Science Foundation, the U.S. Department of Energy, the National
Aeronautics and Space Administration, the Japanese Monbukagakusho, the
Max Planck Society, and the Higher Education Funding Council for
England. The SDSS Web Site is http://www.sdss.org/. The SDSS is managed
by the Astrophysical Research Consortium for the Participating
Institutions. The Participating Institutions are the American Museum of
Natural History, Astrophysical Institute Potsdam, University of Basel,
University of Cambridge, Case Western Reserve University, University of
Chicago, Drexel University, Fermilab, the Institute for Advanced Study,
the Japan Participation Group, Johns Hopkins University, the Joint
Institute for Nuclear Astrophysics, the Kavli Institute for Particle
Astrophysics and Cosmology, the Korean Scientist Group, the Chinese
Academy of Sciences (LAMOST), Los Alamos National Laboratory, the
Max-Planck-Institute for Astronomy (MPIA), the Max-Planck-Institute for
Astrophysics (MPA), New Mexico State University, Ohio State University,
University of Pittsburgh, University of Portsmouth, Princeton
University, the United States Naval Observatory, and the University of
Washington.

{\it Facilities:} \facility{GBT}, \facility{Sloan}.


\begin{thebibliography}{}

\bibitem[Abazajian et al.(2009)]{2009ApJS..182..543A} Abazajian, K.~N., 
Adelman-McCarthy, J.~K., Ag{\"u}eros, M.~A., et al.\ 2009, \apjs, 182, 543 


\bibitem[Allen 
\& Goss(1979)]{1979A&AS...36..135A} Allen, R.~J., \& Goss, W.~M.\ 1979, \aaps, 36, 135 


\bibitem[Allen et 
al.(1978)]{1978A&A....64..359A} Allen, R.~J., van der Hulst, J.~M., Goss, W.~M., \& Huchtmeier, W.\ 1978, \aap, 64, 359 


\bibitem[Babul 
\& Rees(1992)]{1992MNRAS.255..346B} Babul, A., \& Rees, M.~J.\ 1992, \mnras, 255, 346 


\bibitem[Barnes(2011)]{2011MNRAS.413.2860B} Barnes, J.~E.\ 2011, \mnras, 
413, 2860 


\bibitem[Beale 
\& Davies(1969)]{1969Natur.221..531B} Beale, J.~S., \& Davies, R.~D.\ 1969, \nat, 221, 531 


\bibitem[Benson et al.(2002)]{2002MNRAS.333..156B} Benson, A.~J., Lacey, 
C.~G., Baugh, C.~M., Cole, S., \& Frenk, C.~S.\ 2002, \mnras, 333, 156 


\bibitem[Bosma et 
al.(1981)]{1981A&A....93..106B} Bosma, A., Goss, W.~M., \& Allen, R.~J.\ 1981, \aap, 93, 106 


\bibitem[Bothun et al.(1997)]{1997PASP..109..745B} Bothun, G., Impey, C., 
\& McGaugh, S.\ 1997, \pasp, 109, 745 


\bibitem[Bovill 
\& Ricotti(2011)]{2011ApJ...741...18B} Bovill, M.~S., \& Ricotti, M.\ 2011, \apj, 741, 18 


\bibitem[Cannon et al.(2011)]{2011ApJ...739L..22C} Cannon, J.~M., 
Giovanelli, R., Haynes, M.~P., et al.\ 2011, \apjl, 739, L22 


\bibitem[Chynoweth et al.(2011)]{2011AJ....142..137C} Chynoweth, K.~M., 
Holley-Bockelmann, K., Polisensky, E., 
\& Langston, G.~I.\ 2011, \aj, 142, 137 


\bibitem[Chynoweth et al.(2011)]{2011AJ....141....9C} Chynoweth, K.~M., 
Langston, G.~I., \& Holley-Bockelmann, K.\ 2011, \aj, 141, 9 


\bibitem[Chynoweth et al.(2009)]{2009AJ....138..287C} Chynoweth, K.~M., 
Langston, G.~I., Holley-Bockelmann, K., 
\& Lockman, F.~J.\ 2009, \aj, 138, 287 


\bibitem[Chynoweth et al.(2008)]{2008AJ....135.1983C} Chynoweth, K.~M., 
Langston, G.~I., Yun, M.~S., et al.\ 2008, \aj, 135, 1983 


\bibitem[Combes(1991)]{1991A&A...243..109C} Combes, F.\ 1991, \aap, 243, 109 


\bibitem[de Blok et 
al.(2002)]{2002A&A...382...43D} de Blok, W.~J.~G., Zwaan, M.~A., Dijkstra, M., Briggs, F.~H., \& Freeman, K.~C.\ 2002, \aap, 382, 43 


\bibitem[Dekel 
\& Birnboim(2006)]{2006MNRAS.368....2D} Dekel, A., \& Birnboim, Y.\ 2006, \mnras, 368, 2 


\bibitem[Efstathiou(1992)]{1992MNRAS.256P..43E} Efstathiou, G.\ 1992, 
\mnras, 256, 43P 


\bibitem[Freeland 
\& Wilcots(2011)]{2011ApJ...738..145F} Freeland, E., \& Wilcots, E.\ 2011, \apj, 738, 145 


\bibitem[Gallagher et al.(1998)]{1998AJ....115.1869G} Gallagher, J.~S., 
Tolstoy, E., Dohm-Palmer, R.~C., et al.\ 1998, \aj, 115, 1869 


\bibitem[Gooch(1996)]{1996ASPC..101...80G} Gooch, R.\ 1996, Astronomical 
Data Analysis Software and Systems V, 101, 80 


\bibitem[Gunn 
\& Gott(1972)]{1972ApJ...176....1G} Gunn, J.~E., \& Gott, J.~R., III 1972, \apj, 176, 1 


\bibitem[Haynes et al.(1979)]{1979ApJ...229...83H} Haynes, M.~P., 
Giovanelli, R., \& Roberts, M.~S.\ 1979, \apj, 229, 83 


\bibitem[Haynes et al.(2011)]{2011AJ....142..170H} Haynes, M.~P., 
Giovanelli, R., Martin, A.~M., et al.\ 2011, \aj, 142, 170 


\bibitem[Hibbard 
\& Mihos(1995)]{1995AJ....110..140H} Hibbard, J.~E., \& Mihos, J.~C.\ 1995, \aj, 110, 140 


\bibitem[Hibbard et al.(2001)]{2001ASPC..240..657H} Hibbard, J.~E., van 
Gorkom, J.~H., Rupen, M.~P., 
\& Schiminovich, D.\ 2001, Gas and Galaxy Evolution, 240, 657 


\bibitem[Huang et al.(2012)]{2012AJ....143..133H} Huang, S., Haynes, M.~P., 
Giovanelli, R., et al.\ 2012, \aj, 143, 133 


\bibitem[Huchtmeier 
\& Witzel(1979)]{1979A&A....74..138H} Huchtmeier, W.~K., \& Witzel, A.\ 1979, \aap, 74, 138 


\bibitem[Jarrett et al.(2003)]{2003AJ....125..525J} Jarrett, T.~H., 
Chester, T., Cutri, R., Schneider, S.~E., 
\& Huchra, J.~P.\ 2003, \aj, 125, 525 


\bibitem[Kere{\v s} et al.(2009)]{2009MNRAS.395..160K} Kere{\v s}, D., 
Katz, N., Fardal, M., Dav{\'e}, R., 
\& Weinberg, D.~H.\ 2009, \mnras, 395, 160 


\bibitem[Kere{\v s} et al.(2005)]{2005MNRAS.363....2K} Kere{\v s}, D., 
Katz, N., Weinberg, D.~H., \& Dav{\'e}, R.\ 2005, \mnras, 363, 2 


\bibitem[Kniazev et al.(2009)]{2009MNRAS.400.2054K} Kniazev, A.~Y., Brosch, 
N., Hoffman, G.~L., et al.\ 2009, \mnras, 400, 2054 


\bibitem[Kornreich et al.(1998)]{1998AJ....116.2154K} Kornreich, D.~A., 
Haynes, M.~P., \& Lovelace, R.~V.~E.\ 1998, \aj, 116, 2154 


\bibitem[Kornreich et al.(2000)]{2000AJ....120..139K} Kornreich, D.~A., 
Haynes, M.~P., Lovelace, R.~V.~E., \& van Zee, L.\ 2000, \aj, 120, 139 


\bibitem[Kova{\v c} et al.(2009)]{2009MNRAS.400..743K} Kova{\v c}, K., 
Oosterloo, T.~A., \& van der Hulst, J.~M.\ 2009, \mnras, 400, 743 


\bibitem[Langston 
\& Turner(2007)]{2007ApJ...658..455L} Langston, G., \& Turner, B.\ 2007, \apj, 658, 455 


\bibitem[Larson(1972)]{1972Natur.236...21L} Larson, R.~B.\ 1972, \nat, 236, 
21 


\bibitem[Lin 
\& Faber(1983)]{1983ApJ...266L..21L} Lin, D.~N.~C., \& Faber, S.~M.\ 1983, \apjl, 266, L21 


\bibitem[Maller 
\& Bullock(2004)]{2004MNRAS.355..694M} Maller, A.~H., \& Bullock, J.~S.\ 2004, \mnras, 355, 694 


\bibitem[Mangum et 
al.(2007)]{2007A&A...474..679M} Mangum, J.~G., Emerson, D.~T., \& Greisen, E.~W.\ 2007, \aap, 474, 679 


\bibitem[Matheson et al.(2012)]{2012ApJ...754...19M} Matheson, T., Joyce, 
R.~R., Allen, L.~E., et al.\ 2012, \apj, 754, 19 


\bibitem[McGaugh et al.(2000)]{2000ApJ...533L..99M} McGaugh, S.~S., 
Schombert, J.~M., Bothun, G.~D., 
\& de Blok, W.~J.~G.\ 2000, \apjl, 533, L99 


\bibitem[McGaugh(2012)]{2012AJ....143...40M} McGaugh, S.~S.\ 2012, \aj, 
143, 40 


\bibitem[Michel-Dansac et al.(2010)]{2010ApJ...717L.143M} Michel-Dansac, 
L., Duc, P.-A., Bournaud, F., et al.\ 2010, \apjl, 717, L143 


\bibitem[Mihos et al.(2012)]{2012arXiv1211.3095M} Mihos, C., Harding, P., 
Spengler, C., Rudick, C., \& Feldmeier, J.\ 2012, arXiv:1211.3095 


\bibitem[Nichols 
\& Bland-Hawthorn(2011)]{2011ApJ...732...17N} Nichols, M., \& Bland-Hawthorn, J.\ 2011, \apj, 732, 17 


\bibitem[Ott et 
al.(1994)]{1994A&A...284..331O} Ott, M., Witzel, A., Quirrenbach, A., et al.\ 1994, \aap, 284, 331 


\bibitem[Pisano et al.(2007)]{2007ApJ...662..959P} Pisano, D.~J., Barnes, 
D.~G., Gibson, B.~K., et al.\ 2007, \apj, 662, 959 


\bibitem[Pisano et al.(2011)]{2011ApJS..197...28P} Pisano, D.~J., Barnes, 
D.~G., Staveley-Smith, L., et al.\ 2011, \apjs, 197, 28 


\bibitem[Ricotti(2009)]{2009MNRAS.392L..45R} Ricotti, M.\ 2009, \mnras, 
392, L45 


\bibitem[Ricotti 
\& Gnedin(2005)]{2005ApJ...629..259R} Ricotti, M., \& Gnedin, N.~Y.\ 2005, \apj, 629, 259 


\bibitem[Rownd et al.(1994)]{1994AJ....108.1638R} Rownd, B.~K., Dickey, 
J.~M., \& Helou, G.\ 1994, \aj, 108, 1638 


\bibitem[Sancisi et 
al.(2008)]{2008A&ARv..15..189S} Sancisi, R., Fraternali, F., Oosterloo, T., \& van der Hulst, T.\ 2008, \aapr, 15, 189 


\bibitem[Schneider(1985)]{1985ApJ...288L..33S} Schneider, S.\ 1985, \apjl, 
288, L33 


\bibitem[Sommer-Larsen(2006)]{2006ApJ...644L...1S} Sommer-Larsen, J.\ 2006, 
\apjl, 644, L1 


\bibitem[Sternberg et al.(2002)]{2002ApJS..143..419S} Sternberg, A., McKee, 
C.~F., \& Wolfire, M.~G.\ 2002, \apjs, 143, 419 


\bibitem[Stewart et al.(2011)]{2011ApJ...738...39S} Stewart, K.~R., 
Kaufmann, T., Bullock, J.~S., et al.\ 2011, \apj, 738, 39 


\bibitem[Thilker et al.(2007)]{2007ApJS..173..538T} Thilker, D.~A., 
Bianchi, L., Meurer, G., et al.\ 2007, \apjs, 173, 538 


\bibitem[Toomre 
\& Toomre(1972)]{1972ApJ...178..623T} Toomre, A., \& Toomre, J.\ 1972, \apj, 178, 623 


\bibitem[Tully(1988)]{1988ngc..book.....T} Tully, R.~B.\ 1988, Nearby Galaxies Catalog (Cambridge: 
Cambridge Univ. Press)


\bibitem[van der Hulst 
\& Sancisi(1988)]{1988AJ.....95.1354V} van der Hulst, T., \& Sancisi, R.\ 1988, \aj, 95, 1354 


\bibitem[van Zee(2001)]{2001AJ....121.2003V} van Zee, L.\ 2001, \aj, 121, 
2003 


\bibitem[Wakker 
\& van Woerden(1991)]{1991A&A...250..509W} Wakker, B.~P., \& van Woerden, H.\ 1991, \aap, 250, 509 


\bibitem[Waller et al.(1997)]{1997ApJ...481..169W} Waller, W.~H., Bohlin, 
R.~C., Cornett, R.~H., et al.\ 1997, \apj, 481, 169 


\bibitem[Walter et al.(2008)]{2008AJ....136.2563W} Walter, F., Brinks, E., 
de Blok, W.~J.~G., et al.\ 2008, \aj, 136, 2563 


\bibitem[Whiting et al.(2007)]{2007AJ....133..715W} Whiting, A.~B., Hau, 
G.~K.~T., Irwin, M., \& Verdugo, M.\ 2007, \aj, 133, 715 


\bibitem[Yun(1999)]{1999IAUS..186...81Y} Yun, M.~S.\ 1999, in IAU Symp. 186, Galaxy 
Interactions at Low and High Redshift, ed. J. E. Barnes \& D. B. Sanders (Cambridge: 
Cambridge Univ. Press), 81 


\bibitem[Yun et al.(1994)]{1994Natur.372..530Y} Yun, M.~S., Ho, P.~T.~P., 
\& Lo, K.~Y.\ 1994, \nat, 372, 530 




\end{thebibliography}
\end{document}